\documentstyle[preprint,epsf]{jpsj}

\def\runtitle{Correlation Effects in Multi-Band Hubbard Model and Anomalous
Properties of FeSi}
\def\runauthor{Kentaro {\sc Urasaki} and Tetsuro {\sc Saso}}

\title
{
Correlation Effects in Multi-Band Hubbard Model and \\Anomalous
Properties of FeSi
}

\author
{
Kentaro {\sc Urasaki}\footnote{e-mail address: 
kentaro@krishna.th.phy.saitama-u.ac.jp}
and Tetsuro {\sc Saso}\footnote{e-mail address: 
saso@phy.saitama-u.ac.jp}
}

\inst
{
Department of Physics, Faculty of Science, Saitama University, Urawa, 338-8570
}

\recdate
{February 5, 2000}

\abst
{
The two-band Hubbard model with the density of states 
obtained from the band calculation
is applied for FeSi, which is suggested to be
a Kondo insulator or a correlated band insulator.
Using this model, the correlation effects on FeSi 
are investigated in terms of 
the self-consistent second-order perturbation theory combined with the
local approximation. 
The calculated optical conductivity spectrum 
reproduces the experiments by Damascelli {\it et al.} 
semiquantitatively and the specific heat explains the anomalous contribution 
at about 250 K observed in FeSi. Inclusion of the spin fluctuation and the
extension to the case of strong correlation are also discussed.
}

\kword
{two-band Hubbard model, FeSi, optical conductivity, 
specific heat
}

\begin{document}
\sloppy
\maketitle

In the study of a specific material among the strongly correlated electron
systems, the effect of the band structures often plays a crucial role when one
compares a theoretical calculation to the experiments.  Use of a simple
theoretical model might not capture the salient features of the material.
Development of a theoretical method that is capable of taking proper account
of the realistic features of the material is necessary.  We report our recent
approach to the study of the anomalous properties of FeSi in such direction.

FeSi is well known for more than thirty years and a number of studies from
various aspects have been done,
stimulated by the fascinating physical properties.
The early study by Jaccarino {\it et al.}\cite{Jaccarino67}
showed that
the susceptibility is much enhanced over the value expected from the band
paramagnetism at finite temperatures and has a broad peak at about 
500 K.
It was also reported that the specific heat seems 
to have an anomalous enhancement at about 250 K.
These behaviors were explained by a band model with an energy gap, but
unphysically narrow bands were necessary, so that
this difficulty has attracted interests of many researchers.
From the conductivity measurements, 
FeSi is an insulator at low temperatures but shows metallic 
behavior at room temperature. 
To explain these unusual properties of FeSi, 
several theoretical approaches 
have been proposed, 
but the most successful one is the spin fluctuation scenario 
by Takahashi and Moriya.\cite{Takahashi79}
It explains the anomalous magnetic property of FeSi 
and their idea of the thermally induced magnetic moment
was confirmed by the neutron scattering experiment.\cite{Shirane87} 

The recent optical studies\cite{Schlesinger93,Ohta94,Damascelli97,Paschen97},
however, revealed 
the unusual properties of FeSi again.
Schlesinger {\it et al.} reported 
that the gap of about 60 meV ($\sim$700 K) opened 
at low temperatures 
is filled and almost closed at room temperature (about 250$\sim$300 K), 
which they attributed to the correlation effect. 
The following experiments also reported the evidence of 
the correlation effects at low
temperatures.\cite{Saitoh95,Chernikov97,DiTusa98,Fath98} 
In these contexts, Aeppli and Fisk\cite{Aeppli92} suggested that FeSi can be viewed as a 
Kondo insulator or a strongly correlated insulator.

Kondo insulators have been found in the f-electron systems 
and typical examples are YbB$_{12}$\cite{Kasaya85} 
and Ce$_3$Bi$_4$Pt$_3$\cite{Hundley90} and so on.
They have correlated f-bands and small energy gaps at low temperatures. 
Although there are many similarities among FeSi and these materials, 
the correlation in FeSi may not be so strong. 
However, the same physics can be recognized both in 
FeSi and Kondo insulators, 
if one reexamines the experimental data carefully.
From this aspect, Fu and Doniach\cite{Fu95} proposed an 
extended Hubbard model with 
two mixed conduction bands, which is based on their band calculation\cite{Fu94} 
for FeSi, 
and confirmed the importance of the 
correlation effects in physical quantities.
Their calculation, however, seems to include some errors about the 
treatment of the self-energies. 
Therefore, we reinvestigated this model carefully and 
calculated the correlation effects in more correct way,\cite{Urasaki98} 
and confirmed that the correlation effects do play important roles, 
but the shape of the spectrum in the optical conductivity 
did not coincide with the experimental data, 
because of the use of the too simple model Hamiltonian. 

Therefore in the present report, we use an extended two-band Hubbard model 
with the density of states obtained from the band calculation, 
and attempt to explain the low temperature anomalies of FeSi 
observed in the optical conductivity\cite{Urasaki99} and the specific heat 
consistently.

The band calculations\cite{Mattheiss93,Jarlborg95,Galakhov95,Kulatov97,Yamada99} 
for FeSi predict that the ground state 
is a band insulator and a recent calculation\cite{Yamada99} reproduces the 
 gap size close to the observed one. 
Therefore, we start from the band insulator model, 
which consists of two Hubbard bands for d-electrons 
as follows. 
\begin{eqnarray}
H=
& &\sum_{ij\sigma} 
(t^1_{ij}c_{i1\sigma}^\dagger c_{j1\sigma} +t^2_{ij}c_{2i\sigma}^\dagger 
c_{2j\sigma})\cr 
&+&U\sum_{i}(n_{i1\uparrow}n_{i1\downarrow}+ 
n_{i2\uparrow}n_{i2\downarrow})\cr 
&+&U_2\sum_{i}(n_{i1\uparrow}n_{i2\downarrow}+ 
n_{i2\uparrow}n_{i1\downarrow})\cr 
&+&U_3\sum_{i}(n_{i1\uparrow}n_{i2\uparrow}+ 
n_{i2\downarrow}n_{i1\downarrow})\cr 
&-&J\sum_{i}(c_{i1\uparrow}^\dagger c_{i1\downarrow} 
c_{i2\downarrow}^\dagger c_{i2\uparrow} + 
c_{i2\uparrow}^\dagger c_{i2\downarrow} 
c_{i1\downarrow}^\dagger c_{i1\uparrow} ), 
\end{eqnarray} 
where the $c^\dagger_{ia\sigma}(c_{ia\sigma})$ creates (destroys) 
an electron on site $i$ in band $a=$1, 2 with spin $\sigma$. 
The tight binding parameters $t^a_{ij}$ should be fitted 
to the band calculation and $U$, $U_2$, $U_3$ and $J$ 
denote the Coulomb and exchange interactions. 

Since one can expect that the optical 
conductivity spectrum reflects the structure of 
the quasi-particle density of states (DOS) of a system, 
we use the DOS obtained from the 
band calculation for FeSi by Yamada {\it et al.}\cite{Yamada99} 
for the initial DOS 
so as to enable detailed comparison with the experiment.

Furthermore, we start from the following general 
expression of the current operator, 
\begin{eqnarray}
j=e\sum_{\sigma,{\bf k}}\sum_{mm^\prime}
v^{mm^\prime}_{\bf k}c^\dagger_{m{\bf k}}
c_{m^\prime{\bf k}},
\end{eqnarray}
where $m$ denotes the band indices  and 
derive the convenient expression for the optical conductivity. 
For simplicity, 
we set the intra- and interband contributions to be equal 
($v^{mm^\prime}_{\bf k}=v_{\bf k}$). 
Moreover, we assume that the momentum 
conservation is violated in real systems by some 
defects and phonon-assisted transitions. 
Therefore, using the linear response theory, we consider  
the current-current correlation function as below, 
\begin{eqnarray}\label{eq:cfunc2}
K(i\omega_n)&=&
\int^\beta_0d\tau e^{i\omega_n \tau}\sum_{mm^\prime}
\sum_{{\bf kk^\prime}\sigma\sigma^\prime}
v_{\bf k}v_{\bf k^\prime}\cr
&&\!\!\!\times
<T_\tau c^\dagger_{m{\bf k}\sigma}(\tau)c_{m{\bf k}\sigma}(\tau)
c^\dagger_{m^\prime{\bf k^\prime}\sigma^\prime}(0)
c_{m^\prime{\bf k^\prime}\sigma^\prime}(0)>\cr
&&\cr
&&\!\!\!
\!\!\!
\!\!\!
\!\!\!\!\!\!
\!\!\!
\!\!\!
\!\!\!
\simeq-\frac{1}{\beta}\sum_{mm^\prime}
\sum_{l}\sum_{{\bf kk^\prime}\sigma}v_{\bf k}v_{\bf k^\prime}
{\cal G}^m_{{\bf k}\sigma}(i\nu_l)
{\cal G}^{m^\prime}_{{\bf k}^\prime\sigma}(i\nu_l+i\omega_n)\cr
&&\cr&&
\!\!\!
\!\!\!
\!\!\!
\!\!\!
\!\!\!
\!\!\!
\!\!\!
\!\!\!
\times[\delta_{\bf kk^\prime}
+\Gamma_{\bf kk^\prime}^{mm^\prime\sigma}(i\nu_l;i\omega_n)
{\cal G}^m_{{\bf k^\prime}\sigma}(i\nu_l)
{\cal G}^{m^\prime}_{{\bf k}\sigma}(i\nu_l+i\omega_n)
],
\end{eqnarray}
where $\Gamma_{\bf kk^\prime}^{mm^\prime\sigma}(i\nu_l;i\omega_n)$
denotes the vertex function, 
and set $[\dots]$ constant. 
For the present case, this leads to the following expression for the 
optical conductivity,
\begin{eqnarray}\label{eq:j-dos}
\sigma(\omega,T)&=&\frac{\pi(ev)^2}{\hbar}\sum_\sigma
\int^\infty_{-\infty}d\nu 
\frac{f(\nu)-f(\nu+\omega)}{\omega}\cr
&& \hspace{-1cm}\times 
[\rho^\sigma_{1}(\nu)+\rho^\sigma_{2}(\nu)]
[\rho^\sigma_{1}(\nu+\omega)+\rho^\sigma_{2}(\nu+\omega)
], 
\end{eqnarray}
where $\rho^\sigma_{a}(\nu)$ denotes the DOS for the band $a$. 
This joint-DOS-like form for the optical conductivity is 
simple but convenient for the present case. 
We set $(ev)^2/\hbar=1$ for simplicity. 

\begin{figure}\vspace{0.5cm} 
\epsfxsize=8cm
\centerline{\epsfbox{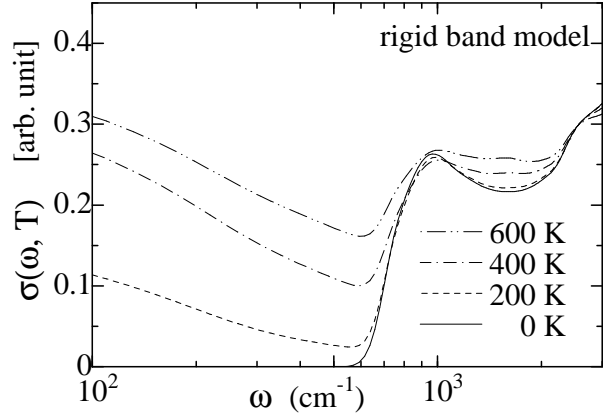}}
\caption{Fig. 1 Calculated optical conductivity 
within the Hartree-Fock approximation or a 
rigid band model.}
\label{fig:1}
\end{figure}
Firstly, we show the optical conductivity obtained 
from the Hartree-Fock approximation (HFA) or a rigid band model in Fig. 1. 
The used DOS is displayed in Fig. 2 by the solid line for $T=0$.
The DOS is independent of the temperature within HFA. 
At 0 K, 
only the interband contribution survives and reproduces 
the shape of the spectrum of the experiment at 4 K in Fig. 3. 
Therefore, 
the band calculation by Yamada {\it et al.}\cite{Yamada99} 
seems to 
give a good result about the whole structure of the DOS at $T=0$ 
but with a slightly smaller gap size
(see the comparison with the experiment below). 
Within the rigid band model, however, since the gap is filled only with 
the intraband (Drude) contribution, 
the temperature variation is monotonous and the spectrum does not become flat  
at a temperature of the order of the gap size.
This disagreement was shown by Fu {\it et al.} first. 
Ohta {\it et al.}\cite{Ohta94} also calculated the optical conductivity 
in the joint-DOS form from their band calculation, but 
the flat part of the optical conductivity spectrum 
within the gap could not be reproduced. 
Therefore, the rigid band model is not sufficient to explain the experiments. 

Next, we investigate the correlation effect in 
the low energy and low temperature region of this model.
Therefore we calculate the correlation effect 
by the self-consistent second-order perturbation theory (SCSOPT)
combined with the local approximation.
The second-order self-energies are given by 
\begin{eqnarray}\label{eq:sigma}
\Sigma_1^{(2)\sigma}(\omega)&=&
\int\!\!\!\int\!\!\!\int^\infty_{-\infty}
d\varepsilon_1 d\varepsilon_2 d\varepsilon_3 \cr\cr
&&[U^2\rho_1^{-\sigma}(\varepsilon_1)
\rho_1^{\sigma}(\varepsilon_2)
\rho_1^{-\sigma}(\varepsilon_3) \cr\cr
&&+U_2^2\rho_2^{-\sigma}(\varepsilon_1)
\rho_1^{\sigma}(\varepsilon_2)
\rho_2^{-\sigma}(\varepsilon_3) \cr\cr
&&+U_3^2\rho_2^{\sigma}(\varepsilon_1)
\rho_1^{\sigma}(\varepsilon_2)
\rho_2^{\sigma}(\varepsilon_3) \cr\cr
&&+J^2\rho_2^{-\sigma}(\varepsilon_1)
\rho_2^{\sigma}(\varepsilon_2)
\rho_1^{-\sigma}(\varepsilon_3) 
]\cr\cr
\times&&
\hspace{-5mm}\frac{f(-\varepsilon_1)f(\varepsilon_2)f(\varepsilon_3)
+f(\varepsilon_1)f(-\varepsilon_2)f(-\varepsilon_3)}
{\omega+\varepsilon_1-\varepsilon_2-\varepsilon_3+{\rm i}\delta},\cr
\Sigma_2^{(2)\sigma}(\omega)&=&(1\leftrightarrow 2),
\end{eqnarray}
where $\rho_a^\sigma(\omega)=
-(1/\pi){\rm Im}G_a^{\sigma}(\omega+{\rm i}\delta)$ and
\begin{eqnarray}\label{eq:g}
G_a^\sigma(\omega)
&&=\frac{1}{N}\sum_{\bf k}G_a^{\sigma}({\bf k},\omega)\cr
&&=\int^{\infty}_{-\infty}d\varepsilon\rho^{0\sigma}_a(\varepsilon)
\frac{1}{\omega-\varepsilon-\Sigma^{(2)\sigma}_a(\omega)}.
\end{eqnarray}
Here, $N$ is the number of sites, $f(\varepsilon)$ the Fermi function 
and $\rho^{0\sigma}_a(\varepsilon)$ 
the DOS of band $a$ for the non-interacting case. 
To make numerical calculation easy, 
we take $\delta$ finite ($\delta=10^{-7}$) in eq. (\ref{eq:g}) and 
convert these equations with the transformations\cite{MullerHartmann89}
\begin{eqnarray}
A_a^\sigma(\tau)&=&\int^\infty_{-\infty}d\epsilon
e^{-{\rm i}\tau\varepsilon}\rho_a^\sigma(\varepsilon)f(\varepsilon),\cr
B_a^\sigma(\tau)&=&\int^\infty_{-\infty}d\epsilon
e^{-{\rm i}\tau\varepsilon}\rho_a^\sigma(\varepsilon)f(-\varepsilon).
\end{eqnarray}
These equations have to be solved 
self-consistently. 
In this paper, we set $U_2=U-J$ and $U_3=U-2J$ 
in order to reduce the number of parameters.
In this case, the Hamiltonian is rotationally invariant 
in spin and real spaces if the two bands are degenerate.\cite{Parmenter73} 

In the following results, $U=0.5$ eV and $J=0.35U$ are chosen 
so as to reproduce the shape and the 
temperature dependence of the optical conductivity spectrum. 
The solid line for $T=0$ in Fig. 2 indicates the initial DOS at 0 K, 
and the correlation effect is absent except the Hartree-Fock contribution 
since the band 1 is filled and the band 2 is empty. 

Note that the gap in the DOS is widened by 16 $\%$ so as to 
reproduce the shape of the spectrum of 
the optical conductivity at 4 K in the experiment, which does not 
change the essence of the following result.  
Then, the gap size ($E_g$) of $75$ meV is obtained 
if the steepest parts of the DOS at 
the both sides of the gap are extrapolated and the tails are neglected. 
(If we regard the gap as the region inside the tails of the gap edge, 
we obtain 60 meV.) 
The band 1 and 2 in our Hamiltonian correspond to 
the upper and lower part of the DOS 
with respect to the Fermi level ($E_F=0$) as is seen in Fig. 2, 
where we introduce a cut off for each band 
so as to include one state per spin in each band. 
Then the band width for the band 1 and 2 are about 0.56 eV and about 
0.85 eV, respectively. 
Although the DOS is asymmetric, 
the chemical potential is fixed at $\omega=0$ 
and assumed to be temperature independent. 
One can see in Fig. 2 that the correlation is introduced  at finite $T$ 
through the thermally excited electrons and holes 
and the gap existing at 0 K is almost filled up 
at the temperature of the order of its size, 
which results in the temperature variation of the interband contribution 
of the optical conductivity (see below). 
\begin{figure}\vspace{0.5cm} 
\epsfxsize=8cm
\centerline{\epsfbox{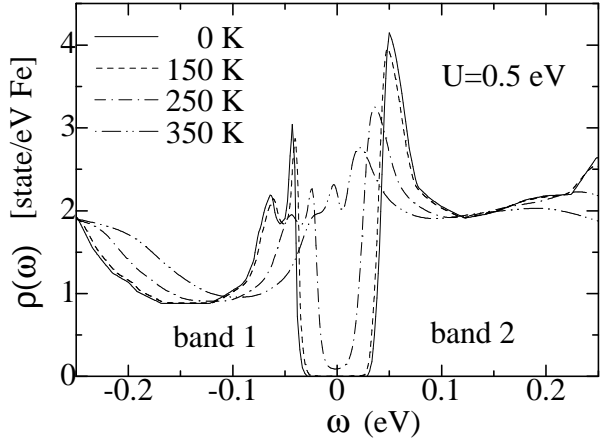}}
\caption{The temperature dependence of the quasi-particle DOS 
and the initial DOS obtained from the band calculation(Ref. 26)
at $T=0$. At finite $T$, the DOS is strongly temperature dependent 
due to the correlation effects. 
}
\label{fig:2}
\end{figure}

\begin{figure}
\epsfxsize=8cm
\centerline{\epsfbox{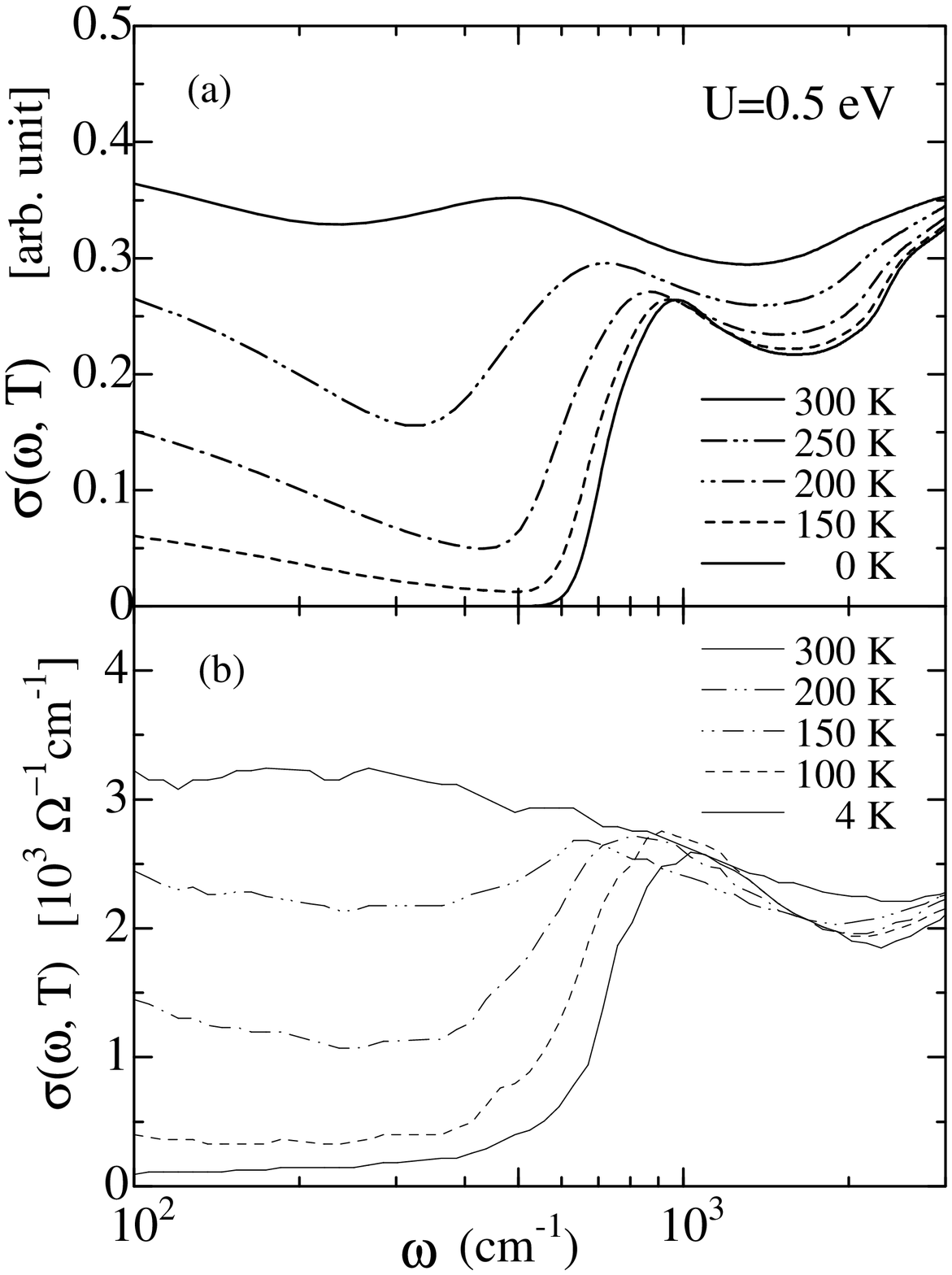}}
\caption{(a)The temperature dependence of the optical conductivity 
calculated with the eq. (\ref{eq:j-dos}). 
(b)The experimental data from Ref. 11. 
The peaks due to phonons observed in the gap are omitted. }
\label{fig:3}
\end{figure}
In Fig. 3(a), the temperature variation of the optical conductivity 
calculated from the temperature-dependent DOS in Fig. 2 is shown. 
In our calculation (Fig. 3(a)), 
the gap is almost filled up at 300 K as well as 
the rapid increase in the gap region from 150 to 300 K is seen. 
This is consistent with the experiment (Fig. 3(b)), 
where the gap is filled rapidly from 100 K to 300 K. 
Reflecting the correlation effects, the peak at the gap edge shifts 
to lower frequency 
region, as is seen in the experiment. 
In our calculation, however, 
there are dips between the Drude and the interband contributions in contrast
to the experiment. 
This may be caused by the simplification 
in deriving eq. (\ref{eq:j-dos}). 
However, the almost flat spectrum is obtained at 300 K, 
which comes from the temperature dependence of the 
interband contribution. 

We also calculate the temperature variation of the specific heat 
with the same parameters as in the optical conductivity. 
Starting from the equation of motion,\cite{Fetter71}
we obtain the following expression for the total energy 
per site:
\begin{eqnarray}\label{eq:sh}
E&=&\frac{1}{2}\sum_\sigma\int^\infty_{-\infty}d\omega
f(\omega)\left[ \omega\{\rho^\sigma_{1}(\omega)+\rho^\sigma_{2}(\omega)\} 
\right.\cr
&& \hspace{-5mm} +\frac{1}{N}\sum_{\bf k}\left. \{\varepsilon^1_{\bf 
k}\rho^\sigma_{1}({\bf k},\omega)
+\varepsilon^2_{\bf k}\rho^\sigma_{2}({\bf k},\omega) \} \right],
\end{eqnarray}
\begin{figure}\vspace{0.5cm} 
\epsfxsize=8cm 
\centerline{\epsfbox{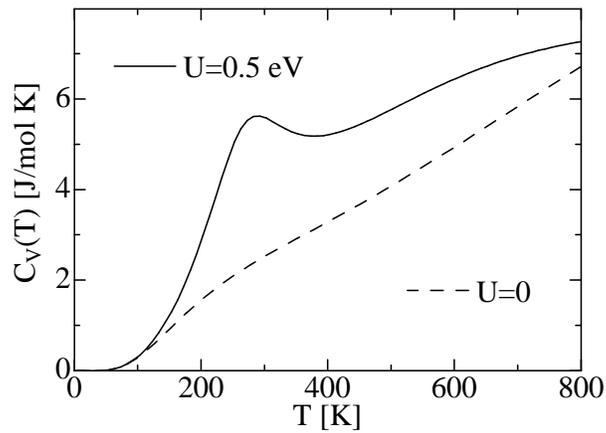}}
\caption{Calculated specific heat using the same parameter as in Fig. 3(a).}
\label{fig:4}
\end{figure}
where $\varepsilon^1_{\bf k}$ ($\varepsilon^2_{\bf k}$) 
is the Fourier transformation of $t^1_{ij}$ ($t^2_{ij}$). 
The specific heat can be calculated from the numerical differentiation 
of the energy as $C_V=(\partial E/\partial T)_V$. 
The difference between the cases with $U=0$ and $0.5$ eV in Fig. 4 
indicates the contribution from the correlation effect, 
which results in a peak of about 4 J/K mol at about 250 K, 
and explains the ``anomalous" 
contribution ($\sim$6 J/K mol) in the specific heat at about 250 K 
reported by Jaccarino {\it et al.}\cite{Jaccarino67}
Note that 
they evaluated the anomaly by subtracting 
the specific heat of CoSi after the 
normal electronic contributions $\gamma_{\rm FeSi}$ 
and $\gamma_{\rm CoSi}$ are removed, respectively. 
In the above calculations, we confirmed that the correlation effect is
essential 
to explain the temperature dependence of the optical conductivity
and the specific heat in FeSi.
At higher temperatures or for magnetic properties, however,
it is also important to take the spin fluctuations
\cite{Takahashi79,Saso99} 
into account. 

The self-consistent renormalization (SCR) theory of spin fluctuations 
has succeeded in describing the itinerant magnetism and 
the quantum critical phenomena 
with a small number of parameters.\cite{Moriya85} 
On the other hand, the dynamical mean field theory (DMFT) 
is one of the most powerful schemes to take account of the 
strong local correlation. 
One of the authors has proposed a new and practical scheme 
that unifies DMFT and SCR in a microscopic way.\cite{Saso99} 
Application of this theory to FeSi may improve the present calculation towards 
the inclusion of the effects of spin fluctuations at finite temperatures and 
the intermediate coupling. 

\section*{Acknowledgements}
The authors would like to thank Professor H. Yamada for providing them 
the details of the band calculation (LMTO-ASA) for FeSi and 
for his useful comments.
This work is supported by Grant-in-Aid for Scientific Research No.11640367
from the Ministry of Education, Science, Sports and Culture.




\end{document}